\documentclass[10pt, journal, compsoc]{IEEEtran}
\ifCLASSOPTIONcompsoc
  \usepackage[nocompress]{cite}
\else
  \usepackage{cite}
\fi
\usepackage{graphicx}
\usepackage{amsmath}
\usepackage{algorithm}
\usepackage{algorithmic}
\ifCLASSOPTIONcompsoc
  \usepackage[caption=false,font=footnotesize,labelfont=sf,textfont=sf]{subfig}
\else
  \usepackage[caption=false,font=footnotesize]{subfig}
\fi
\usepackage{url}

\usepackage{hyperref}
\hypersetup{hidelinks}

\newtheorem{definition}{Definition}
\usepackage{multirow}
\usepackage{enumerate}
\usepackage{booktabs}

\usepackage[OT1]{fontenc} 

\begin{document}
%
\title{A Survey of Constrained Combinatorial Testing}
%
%
%
%

\author{Huayao~Wu,
        Changhai~Nie, 
        Justyna~Petke,
        Yue~Jia,
        and Mark~Harman
\IEEEcompsocitemizethanks{
\IEEEcompsocthanksitem H. Wu and C. Nie are with Department of Computer Science and Technology, Nanjing University, Nanjing, China, 210023.\protect\\
E-mail: \{hywu, changhainie\}@nju.edu.cn
\IEEEcompsocthanksitem J. Petke is with CREST, Computer Science, University College London, London, UK, WC1E 6BT.\protect\\
E-mail: j.petke@ucl.ac.uk
\IEEEcompsocthanksitem Y. Jia and M. Harman are with Facebook Inc., London, UK, W1T 1FB and CREST, Computer Science, University College London, London, UK, WC1E 6BT.\protect\\
E-mail: \{yue.jia, mark.harman\}@ucl.ac.uk}
}

\IEEEtitleabstractindextext{%
\begin{abstract}
Combinatorial Testing (CT) is a potentially powerful testing technique, whereas its failure revealing ability might be dramatically reduced if it fails to handle constraints in an adequate and efficient manner. To ensure the wider applicability of CT in the presence of constrained problem domains, large and diverse efforts have been invested towards the techniques and applications of constrained combinatorial testing.
In this paper, we provide a comprehensive survey of representations, influences, and techniques that pertain to constraints in CT, covering 129 papers published between 1987 and 2018. This survey not only categorises the various constraint handling techniques, but also reviews comparatively less well-studied, yet potentially important, constraint identification and maintenance techniques.
Since real-world programs are usually constrained, this survey can be of interest to researchers and practitioners who are looking to use and study constrained combinatorial testing techniques.
\end{abstract}

}

\maketitle

\IEEEdisplaynontitleabstractindextext

%
\IEEEpeerreviewmaketitle

\IEEEraisesectionheading{\section{Introduction}\label{sec_introduction}}

%
%
%
%
\IEEEPARstart{E}{ver} since its initial concept was sketched in 1980s~\cite{Mandl:1985.1}, Combinatorial Testing (CT) has become an important and indispensable testing technique for the quality assurance of modern software systems~\cite{Kuhn:2004.59}.
Traditionally, CT assumes that the parameters of software under test are independent from each other. As such, the mathematical objects named $\tau$-way covering arrays can be directly used as test suites to systematically examine the interactions between any $\tau$ or fewer parameters~\cite{Nie:2011.238}.
However, in real-world programs, there usually have dependency relationships, i.e., {\em constraints}, between parameters~\cite{Petke:2015.538}. Such constraints may simply indicate that a particular interaction cannot be achieved in the test case (for example, the Linux operating system cannot be combined with the IE browser); they can also be more pernicious as a constraints-violating test case might still be executed but yield results that are difficult to distinguish from a software failure.
As a result, any application of CT that fails to take constraints into account will lead to many invalid, or ineffective, test cases; CT could then be less effective than people would otherwise expect.

In order to promote the successful applications of CT in the widespread constrained problem domains, a lot of techniques have been developed and applied.
Especially, there is a variety of constraint handling techniques that aid in generation of constraints-satisfying test cases, ranging from the manual modifications of test models~\cite{Sherwood:1994.10, Grindal:2006.98}, to the automated strategies that rely on constraint satisfiability solvers~\cite{Cohen:2008.131, Garvin:2010.207} and minimal forbidden tuples~\cite{Yu:2014.461, Yu:2015.482}.
Additionally, techniques that try to automatically infer~\cite{Nakagawa:2015.537, Nakagawa:2015.614}, validate~\cite{Tzoref-Brill:2016.578}, and repair~\cite{Gargantini:2016.615, Gargantini:2017.659} constraints are also proposed in more recent studies.

However, even though the topic of constraints has been discussed in many studies, the issues of adequately handling constraints remains challenging in CT.
In 2014, Khalsa and Labiche~\cite{Khalsa:2014.463} conducted a survey based on 75 test suite generation algorithms and tools in CT. They found that more than half of these studies simply do not implement any constraint handling technique, which makes them inapplicable to many real-world programs.

\begin{figure*}[!t]
\centering
\includegraphics[width=\textwidth]{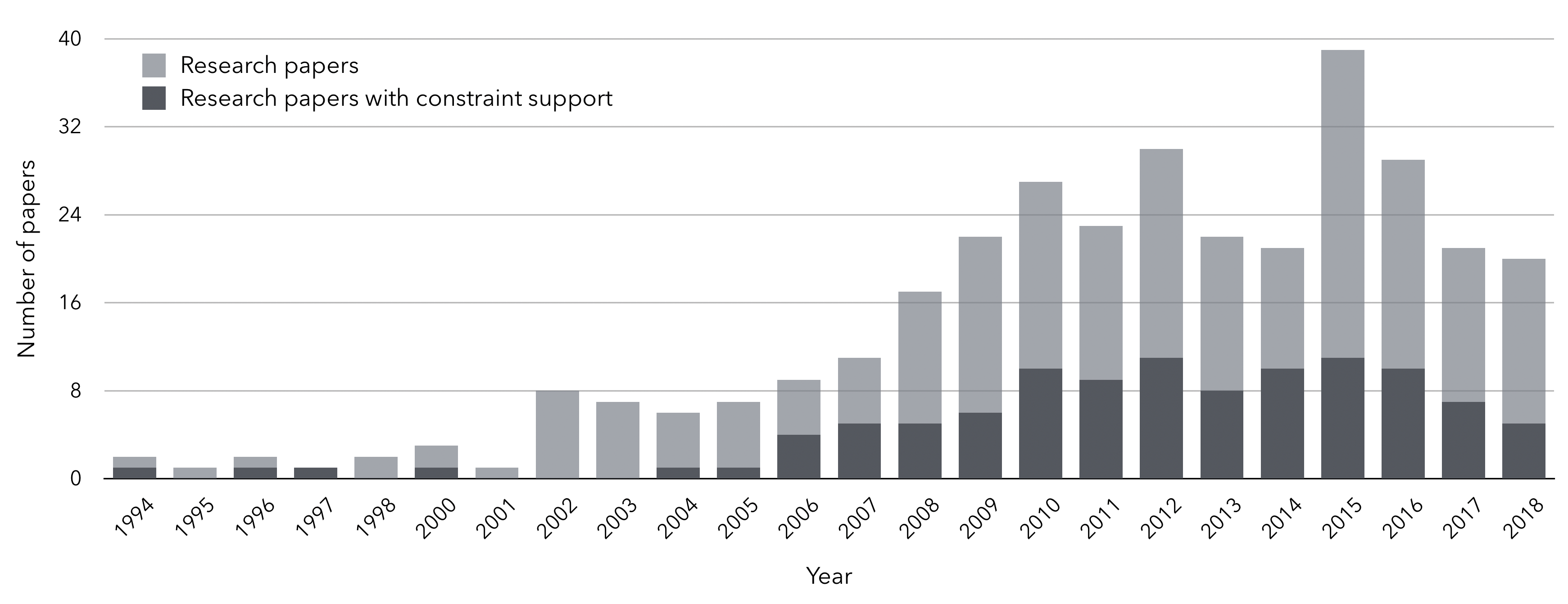}
\caption{Number of research papers on CT test suite generation, and the papers on constraint support in CT by year.}
\label{fig_publication_gen}
\end{figure*}

Moreover, Figure~\ref{fig_publication_gen} further shows the number of research papers on CT test suite generation, and the papers that have taken constraints into account by year, up to 2018.\footnote{These papers are collected from Combinatorial Testing Repository (\url{http://gist.nju.edu.cn/ct_repository}). The processes to construct this repository and to select these papers are explained in Section~\ref{sec_search}.}
We can see that only 32\% of test suite generation studies have incorporated constraint handling techniques. Even in the last four years, where one might suppose that the importance of constraints should have been well-recognised, this proportion remains low, at only 30\%.
Therefore, the importance of constraints, and the need for constrained combinatorial testing, clearly remains a pressing concern for wider adoption of CT in practice.

In this paper, we present a detailed and comprehensive survey of constraints in CT, with the goal to consolidate and distill the large amount of studies in the growing field of constrained combinatorial testing.
To this end, we first collected 129 research papers published from 1987 to 2018. We then overviewed the various representations used to capture constraints, and also the influence of constraints on the successful applications of CT. In particular, we classified the papers into constraint identification, constraint handling, and constraint maintenance, according to their research topics. The available techniques in each category (especially, the constraint handling group) are then extensively reviewed and analysed.

This survey is structured as follows.
Section~\ref{sec_search} describes the process used to find relevant publications. 
Section~\ref{sec_cct} sets out the background on constrained combinatorial testing, including constraint representations and impacts of constraints in CT.
Sections~\ref{sec_identification},~\ref{sec_handling} and~\ref{sec_maintenance} review and analyse pertinent techniques of constraint identification, handling and maintenance, respectively.
Finally, Section~\ref{sec_conclusion} concludes this paper.

\section{Literature Search and Selection}
\label{sec_search}

In this section, we describe the literature search process to construct our combinatorial testing repository, and the process used to select papers on constrained combinatorial testing.

\subsection{Combinatorial Testing Repository}

We constructed the combinatorial testing repository with the aim of providing a full coverage of publications in the literature on combinatorial testing (CT). The primary inclusion criterion is that the paper focuses on research on CT, which should address at least one testing activity, including but not limited to modelling, test suite generation, test suite selection and prioritisation, fault diagnosis, evaluation and applications of CT~\cite{Nie:2011.238}.
In addition, the paper should be presented in English, and should not be purely mathematical studies that solely focus on generating covering arrays by mathematical techniques\footnote{Readers might refer to recent work~\cite{Moura:2016finite, Sarkar:2017upper, Sarkar:2018partial, Kampel:2019algebraic, Colbourn:2019constructions} for more information about such mathematical studies. There is also a survey~\cite{Torres-Jimenez:2013.394} that covers a variety of mathematical based covering array generation methods.}, 
as such studies are not concerned with CT per se, but with the general problem of covering arrays.

The literature search was conducted on six major online library search engines: IEEE Xplore, ACM Digital Library, Elsevier ScienceDirect, Springer Link, Wiley Online Library and DBLP. We searched for related publications using queries ``combinatorial testing'', ``combinatorial interaction testing'', ``pairwise testing'', ``t-way testing'' and ``covering array''. We then collected unique publications, where each of them is manually filtered based on our inclusion and exclusion criterion: we first read the title of each publication and removed those that are clearly irrelevant. For the potentially relevant publications, we further inspected their abstracts (and the full bodies when necessary) to decide whether to include them.
Finally, for the publications that are published in the major conference proceedings and journals of software engineering\footnote{These proceedings and journals are adapted from a previous study~\cite{Karanatsiou:2019}, indicating high quality venues for combinatorial testing publications. We additionally included two specific workshops (IWCT and CTA) on combinatorial testing.} (as shown in Table~\ref{table_venue}), we went through their references to identify new publications, in order to mitigate the risk of omitting relevant studies (snowballing).

\begin{table}[!t]
\renewcommand{\arraystretch}{1.3}
\scriptsize
\caption{Selected Conference Proceedings and Journals.}
\label{table_venue}
\centering
\begin{tabular}{p{6.8cm} l}
\toprule
Venue & Abbr. \\
\midrule
International Conference on Software Engineering & ICSE \\
European Software Engineering Conference and Symposium on the Foundations of Software Engineering  & ESEC/FSE \\
International Conference on Automated Software Engineering & ASE \\
International Symposium on Software Testing and Analysis & ISSTA \\
International Symposium on Software Reliability Engineering & ISSRE \\
International Conference on Software Testing, Verification and Validation & ICST \\
International Workshop on Combination Testing & IWCT \\
International Workshop on Combinatorial Testing and its Applications & CTA \\
IEEE Transactions on Software Engineering & TSE \\
ACM Transactions on Software Engineering and Methodology & TOSEM \\
IEEE Software & IEEE SW \\
Empirical Software Engineering & EMSE \\
Information and Software Technology & IST \\
Journal of Systems and Software & JSS \\
Software Testing, Verification and Reliability & STVR \\
\bottomrule
\end{tabular}
\end{table}

For every included publication, our combinatorial testing repository not only contains its meta-data (such as author, title, conference proceeding or journal information, and publication year), but also classifies it into one of the six research fields (these fields are introduced in a previous survey that reviews works on CT overall~\cite{Nie:2011.238}):
\begin{enumerate}
	\item {\bf Modelling}. Studies on identifying parameters, values, and the interactions between parameters of the software under test.
	\item {\bf Generation}. Studies on generating the smallest CT test suite.
	\item {\bf Optimisation}. Studies on improving the test suite by prioritisation, minimisation and selection techniques.
	\item {\bf Fault Diagnosis}. Studies on locating the concrete failure-causing interactions.
	\item {\bf Evaluation}. Studies on measuring the effectiveness of CT and comparing CT with other testing methods.
	\item {\bf Application}. Studies on applying, improving, and popularising CT and its procedures in the real-world.
\end{enumerate}
Note that some publications may focus on multiple fields at the same time, but we nevertheless assign each publication into one field based on its main contribution. As a result, the distribution on the whole can provide a fairly good picture of the current state of research in CT~\cite{Nie:2011.238}.

The combinatorial testing repository was initially available online in late 2015, containing 461 publications at that time. Since then, we repeatedly conducted manual searches on major conferences and journals (as shown in Table~\ref{table_venue}), in order to keep this repository up-to-date. The snowballing process is also performed for the newly identified publications.

Moreover, in order to ensure that all relevant publications are indeed included for this study, we re-conducted the whole literature search process in 2018 to further underpin the combinatorial testing repository. The repository currently contains 764 publications (December 2018). It is publicly available online at \url{http://gist.nju.edu.cn/ct_repository}.
\footnote{Since the contents in the repository might be changed due to further updates, we additionally created a mirror archive that includes the 764 papers used in this study. The complete list of these papers is available online at \url{https://github.com/GIST-NJU/ctrepo-archive}.}

We note that there is another publicly available repository on combinatorial testing\footnote{\url{https://cs.unibg.it/gargantini/citrepo/}}. However, that repository only contains 428 publications, and has not been updated since 2016. In addition, that repository contains publications that are not written in English. Such non-English language publications are excluded in our study. Nevertheless, we have gone through all publications in that repository to make sure that every publication that meets our inclusion criteria is also included in our repository.

\subsection{Paper Selection for This Survey}

The focus of this study is constraint-related research in CT. Hence we selected relevant publications from the combinatorial testing repository. We considered journal articles, conference and workshop papers, technical reports and Ph.D. theses that are published before December 2018. The main inclusion criterion is that the paper is related to constrained combinatorial testing, concerning the impacts and specific techniques when the input space of the software under test is constrained. We exclude papers that simply focus on the application of constrained combinatorial testing, in which existing algorithms or tools are directly applied on particular (constrained) software systems. 

\begin{figure}[!t]
\centering
\includegraphics[width=0.48\textwidth]{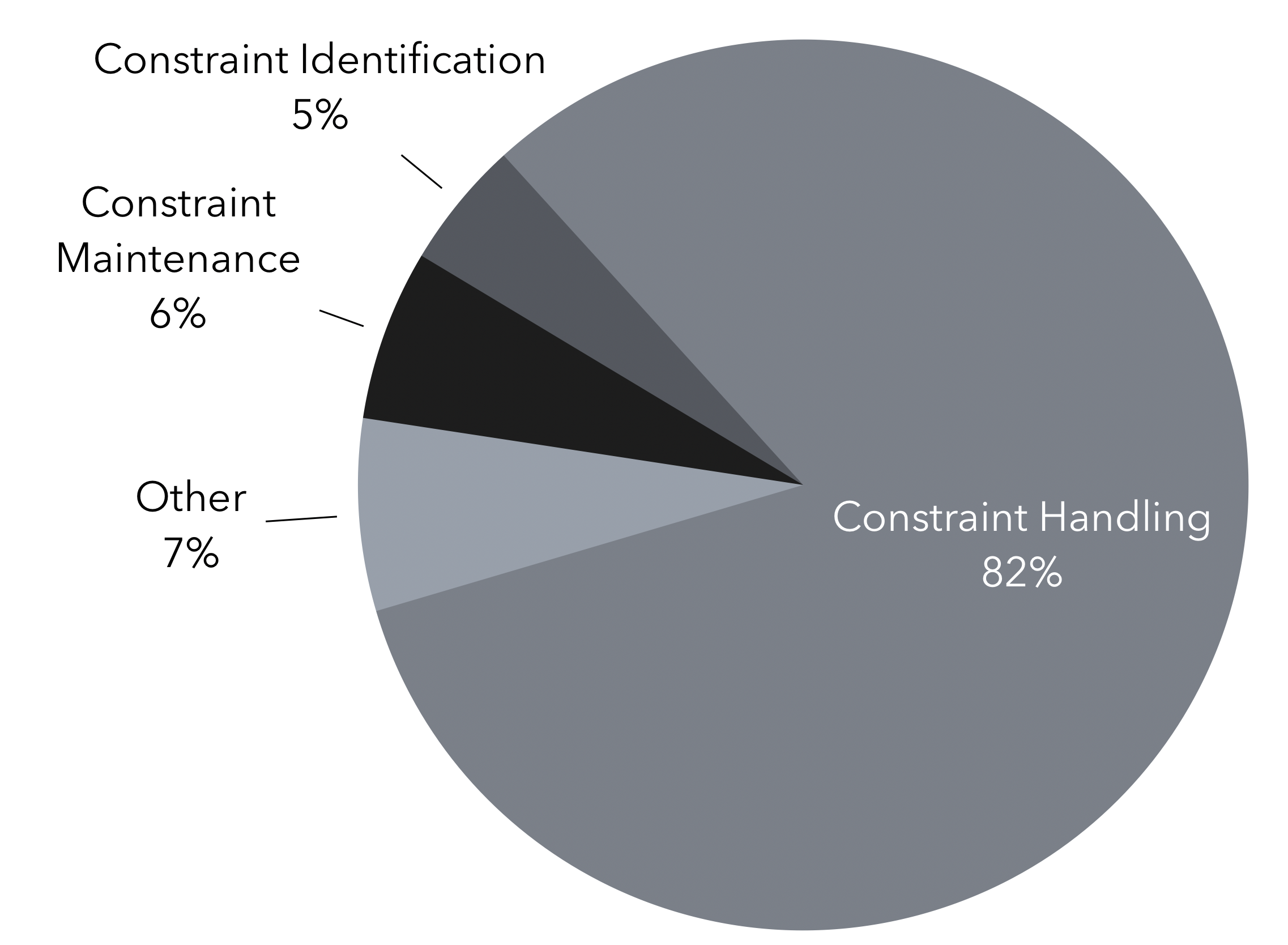}
\caption{Distribution of research topics in constrained combinatorial testing.}
\label{fig_distribution}
\end{figure}

We went through all publications in the combinatorial testing repository against the above inclusion and exclusion criteria. We also applied these criteria on all references in previous SLR paper~\cite{Ahmed:2017} to ensure that every relevant publication is collected (note that this previous study covers papers on applications of constrained combinatorial testing, which are excluded in our study). Finally, we arrived at a total of 129 publications for this survey (published between 1987 and 2018). 

We classified the collected studies into the following three main categories, based on their research topics: 
\begin{enumerate}
\item {\bf Constraint Identification}: Studies on identifying potential constraints of software under test.
\item {\bf Constraint Handling}: Studies on effectively generating constraints-satisfying test suites (i.e., constrained covering arrays).
\item {\bf Constraint Maintenance}: Studies on validating, repairing and evolving constraints in the test model.
\end{enumerate}
The distribution of publications in each category is presented in Figure~\ref{fig_distribution}. It is obvious that constraint handling is the most prominent field (106 papers). This is mainly due to the popularity of test suite generation research in CT~\cite{Nie:2011.238}. There are also a few publications on constraint identification and maintenance; this is the first study in the literature to review such topics.

\section{Constrained Combinatorial Testing}
\label{sec_cct}

We begin our survey with background on constrained combinatorial testing, describing constraint representations in CT and the impact of constraints in CT application.

\subsection{Background}
\label{sec_definition}

Combinatorial testing (also known as combinatorial interaction testing) is a systematic technique that selects combinations of program inputs or features for testing~\cite{Cohen:2010comb}. Such inputs in a Software Under Test (SUT) could represent system configurations, internal or external events, user inputs, etc.

\begin{definition}[Test case~\cite{Nie:2011.238}]
Let $P$ be a set of $n$ parameters $P=\{p_1, p_2, \ldots, p_n\}$ and $V=\{V_1, V_2, \ldots, V_n\}$, where each $V_i$ is a finite set of discrete values, and each parameter $p_i$ can take values from $V_i$ for all $1 \leq i \leq n$.
We call an $n$-tuple ($x_1$, $x_2$, $\ldots$, $x_n$) a test case $t$, where each $x_i$ is a valid instantiation of $p_i$, i.e., $x_i \in V_i$ for $1 \leq i \leq n$. 
\end{definition}

\begin{definition}[$\tau$-way combination~\cite{Nie:2011.238}]
Let $t=(x_1, x_2, \ldots, x_n)$ be a test case over $P=\{p_1, p_2, \ldots, p_n\}$ parameters. A $\tau$-way combination is a combination of any $\tau$ parameters of $t$, where $1 \leq \tau \leq n$. We will use the following notation to represent a $\tau$-way combinations: ($-$, ${x}_{k_1}$, \ldots, ${x}_{k_\tau}$, \ldots) where each of the $\tau$ parameters have fixed values and the other parameters are assigned any valid value from their domain, represented  as ``$-$''.
\end{definition}

Exhaustive testing covers all $n$-way combinations by definition. Such test suites, however, are usually prohibitively large. Instead, CT provides a systematic approach to selecting a subset of all possible inputs, with the aim to only cover every $\tau$-way combination at least once. The concept of CT comes from the fact that if no more than $\tau$ parameters are involved in any failure, then covering all $k$-way combinations ($k \leq \tau$) is effectively equivalent to exhaustive testing. Hence, a mathematical object named a $\tau$-way covering array is used to represent CT test suites.

\begin{definition}[covering array~\cite{Nie:2011.238}]
A $\tau$-way covering array is an $N \times n$ array that satisfies the following properties: (1) each column $i$ ($1 \leq i \leq n$) contains only element from the set $V_i$; and (2) the rows of each $N \times \tau$ sub array cover all $|V_{k_1}| \times |V_{k_2}| \times \ldots \times |V_{k_\tau}|$ combinations of the $\tau$ columns at least once, where $1 \leq \tau \leq n$ and $1 \leq k_1 < \ldots < k_\tau \leq n$.
\end{definition}

A $\tau$-way covering array is often denoted by $CA(N; \tau, v_1^{k_1} v_2^{k_2} \ldots v_m^{k_m})$, where $v_i^{k_i}$ stands for $k_i$ parameters with the same number of $v_i$ values, $\sum k_i = n$. 

The value of $\tau$ is referred to as the {\em covering strength}. Determining this value is a key issue in CT. The empirical observations of Kuhn et al. have demonstrated that most software failures are caused by the interactions of one or two parameters, and the value of $\tau$ is not likely to exceed six~\cite{Kuhn:2004.59}.
Hence, $\tau=2$, or pairwise, is the most widely used choice in practice, which can achieve a good balance between the size of test suite and failure finding effectiveness.

Table~\ref{table_example}, for example, shows a test model for testing font effects in a word processor (this example is adapted from a previous work~\cite{Wu:2018empirical}). This test model has $n=5$ parameters with $|V_1| = |V_3| = 3$ and $|V_2| = |V_4| = |V_5| = 2$.
Instead of exhaustively examining all $3^2 \times 2^3 = 72$ test cases, 2-way CT only requires 9 test cases to cover every 2-way combination at least once. Table~\ref{table_covering} shows such a 2-way covering array, where each row is exactly a test case of the test model. 

\begin{table}[!t]
\renewcommand{\arraystretch}{1.3}
\caption{A test model for `font effect' (adapted from a previous work~\cite{Wu:2018empirical}).}
\label{table_example}
\centering
\begin{tabular}{c c c c c}
\toprule
\multirow{2}{*}{Style} & \multirow{2}{*}{Underline} & Underline & \multirow{2}{*}{Superscript} & \multirow{2}{*}{Subscript} \\
& & Colour & & \\
\midrule
Regular & On   & Red   & On  & On \\
Italic  & Off  & Blue  & Off & Off \\
Bold    &      & Green &     & \\
\bottomrule
\multicolumn{5}{l}{Constraint: {\em Superscript} and {\em Subscript} cannot both be enabled} \\
\end{tabular}
\end{table}

\begin{table}[!t]
\renewcommand{\arraystretch}{1.3}
\caption{A 2-way covering array $CA(9; 2, 3^1 2^1 3^1 2^2)$.}
\label{table_covering}
\centering
\begin{tabular}{c|c c c c c}
\toprule
& \multirow{2}{*}{Style} & \multirow{2}{*}{Underline} & Underline & \multirow{2}{*}{Superscript} & \multirow{2}{*}{Subscript} \\
& & & Colour & & \\\midrule
$t_1$ & Regular & On  & Red   & Off & Off \\ 
$t_2$ & Regular & On  & Blue  & On  & Off \\ 
$t_3$ & Regular & Off & Green & Off & On  \\ 
$t_4$ & Italic  & Off & Red   & On  & On  \\ 
$t_5$ & Italic  & Off & Blue  & Off & Off \\ 
$t_6$ & Italic  & On  & Green & On  & Off \\ 
$t_7$ & Bold    & On  & Red   & On  & On  \\ 
$t_8$ & Bold    & Off & Blue  & On  & On  \\ 
$t_9$ & Bold    & Off & Green & Off & Off \\ 
\bottomrule
\end{tabular}
\end{table}

The discussion so far assumes that every possible $\tau$-way combination is feasible and has the potential to trigger the failure. However, this may be unrealistic due to the constraints between parameter values. Constraints may be introduced because of inconsistencies between hardware components, limitations on possible configurations, or simply design choices~\cite{Cohen:2007.114}. 
For example, one constraint for the model in Table~\ref{table_example} is that `{\em Superscript} and {\em Subscript} cannot both be enabled for the same character'. A test case that violates this constraint is considered invalid. Examples are test cases $t_4$, $t_7$ and $t_8$ in Table~\ref{table_covering}.

In order to incorporate constraints into CT, the definition of a covering array needs to be extended to that of constrained covering array, which can be defined as follows:

\begin{definition}[constrained covering array~\cite{Cohen:2008.131}]
A $\tau$-way constrained covering array with respect to a set of constraints $C$ is an $N \times n$ array, where (1) each column $i$ ($1 \leq i \leq n$) contains only elements from the set $V_i$; (2) each row is $C$-satisfying; and (3) it covers every $C$-satisfying $\tau$-way combination at least once.
\end{definition}

Constraints can be either {\em hard} or {\em soft}. A hard constraint requires that certain parameter combinations cannot appear in any test case, because their existence will prevent the test case from execution. The constraint in Table~\ref{table_example} is an example of a hard constraint.
A soft constraint, on the other hand, is the combination that does not need to be tested, based on the knowledge and experience of testers. It is possible to include test cases that violate soft constraints, but these are undesirable and bring no benefit to test effectiveness. The concept of a soft constraint was initially proposed in 2006~\cite{Bryce:2006.92}, and only few studies have focused on such constraints since then~\cite{Bryce:2006.92, Bryce:2007.121}.

In addition, sometimes a given software system needs to be tested by running a set of selected test cases in a set of selected configurations. In this case, a constraint can be either {\em system-wide} or {\em test case-specific}~\cite{Yilmaz:2012.308}. System-wide constraints determine the valid space from which the configurations are sampled, while test case-specific constraints determine the set of configurations in which a test case could run. These two constraints can be accounted for at the same time by an object named test case-aware covering array~\cite{Yilmaz:2012.308}.

Intuitively, all constraints are explicitly specified in the test model. Sometimes the interactions of a set of constraints may give rise to new constraints. Such a newly introduced constraint is named an {\em implicit constraint}, because usually such implicit constraints are unknown to the tester.
For example, for the test model in Table~\ref{table_example}, we already have one invalid combination ($-$, $-$, $-$, {\em On}, {\em On}). If we add another invalid combination ($Bold$, $-$, $-$, {\em Off}, $-$) as a constraint, then a new invalid 2-way combination ({\em Bold}, $-$, $-$, $-$, {\em On}) is introduced, because we cannot find a test case that covers both this combination and does not cover the two explicitly given invalid combinations.

\subsection{Representations of Constraints}

Constraints in CT can be represented by different forms and their combinations. In this section, we describe four common representations of constraints: {\em forbidden}, {\em implication}, {\em numeric} and {\em shielding}, and another two recently proposed representations of improving the tester's capability of expressing constraints: {\em counter \& value property} and {\em embedded function}.

We focus on abstract representations here, because different algorithms and tools may use different modelling languages. A proposal of a common language to represent CT problems can be found in the \textsc{CitLab} project~\cite{Gargantini:2012.283}.

\subsubsection{Forbidden}

The forbidden, or forbidden tuple approach, is probably the most straightforward constraint representation. Each forbidden constraint is exactly an invalid $k$-way combination, indicating that some parameter values cannot appear together in a test case. For example, the constraint `{\em Superscript} and {\em Subscript} cannot both be enabled' can be represented as a 2-way forbidden tuple, ($-$, $-$, $-$, $On$, $On$), for the test model in Table~\ref{table_example}.

\subsubsection{Implication}

Sometimes the allowable values of some parameters are determined by the value assignments of other parameters. Such a constraint can be represented by an implication relationship, denoted as $p \rightarrow q$ (if $p$ is true, then $q$ must also be true). For example, the constraint `if {\em Style} is set to {\em Bold}, then {\em Superscript} should be enabled' can be represented as $\textit{Style} = \textit{Bold} \rightarrow \textit{Superscript} = \textit{On}$ for the test model in Table~\ref{table_example}.

Forbidden tuple and implication constraints are the most common constraint representations in CT. They can be easily transformed into each other.
To transform a forbidden tuple constraint containing $k$ parameters into implication constraints, we can construct $k$ implication constraints, each of which uses one $(k-1)$-way combination of this forbidden tuple as the premise of the implication. For example, ($-$, $-$, $-$, $On$, $On$) in Table~\ref{table_example} can be transformed into two implication constraints: $\textit{Superscript} = \textit{On} \rightarrow \textit{Subscript} = \mathit{Off}$ and $\textit{Subscript} = \textit{On} \rightarrow \textit{Superscript} = \mathit{Off}$. 
To transform an implication constraint into forbidden tuple constraints, we can directly negate this implication. For example, an implication constraint, $\textit{Style} = \textit{Bold} \rightarrow \textit{Underline Colour} = \textit{Red}$ in Table~\ref{table_example}, can be transformed into two forbidden tuple constraints: ($Bold$, $-$, $Blue$, $-$, $-$) and ($Bold$, $-$, $Green$, $-$, $-$).

\subsubsection{Numeric}

Sometimes some parameter values should satisfy a given arithmetical relationship~\cite{Kruse:2012.324}. This can be represented by a numeric constraint, denoted as $(f, P_f)$, where $P_f \subset P$ is a set of parameters and $f$ is the relationship on $P_f$ that must be satisfied. For example, given a test model with parameter set $P = \{a, b, c\}$ where each parameter can take values from $\{0, 1, 2\}$, a numeric constraint may be $(f, P)$ where $f: a + b + c \geq 3$.

Numeric constraints can also be transformed into forbidden tuple constraints. This can be done by examining the validity of every combination of all parameters in $P_f$. However, this may result in a large number of forbidden tuples. For example, for the above test model, transforming the numeric constraint $(a + b + c \geq 3, \{a, b, c\})$ will result in 10 forbidden tuples: $(0, 0, 0)$, $(0, 0, 1)$, $(0, 0, 2)$, $\cdots$, $(2, 0, 0)$.

\subsubsection{Shielding}

Generally CT assumes that all parameters are always active, namely each parameter must take one and exactly one value. In some circumstances, however, the existence of a combination may invalidate other parameter choices, so that these parameters cannot take unconstrained values~\cite{Chen:2010.203}.
Such a constraint is named a shielding constraint, denoted as $a/P_a$, where $a$ is a $k$-way combination and $P_a$ is a set of parameters that will be invalid when $a$ is assigned. The combination $a$ is called a shielding combination, while the parameters in $P_a$ are called dependent parameters of $a$.
For example, for the test model in Table~\ref{table_example}, there is one shielding constraint `if {\em Underline} takes value {\em Off}, then the parameter {\em Underline Colour} will be disabled', namely ($-$, $\mathit{Off}$, $-$, $-$, $-$)$/$\{{\em Underline Colour}\}. 

Shielding constraints need special consideration as they cannot be directly transformed into other constraints. This is because the dependent parameters will not take values when they are disabled, which contradicts the requirement that each parameter must take one value in a test case. An alternative solution is to add a void value `$\#$' to the value set of all dependent parameters, and add constraints to ensure that if $a$ appears then all parameters in $P_a$ should take $\#$, and if any parameter in $P_a$ takes $\#$ then $a$ should be assigned.
For example, to transform the shielding constraint ($-$, $\mathit{Off}$, $-$, $-$, $-$)$/$\{{\em Underline Colour}\}, we can add a value $\#$ for parameter {\em Underline Colour} and add two implication constraints: $\textit{Underline} = \textit{Off} \rightarrow \textit{Underline Colour} = \#$ and $\textit{Underline Colour} = \# \rightarrow \textit{Underline} = \textit{Off}$.

\subsubsection{Counter \& Value Property}
In addition to the above four common representations of constraints, Segall et al.~\cite{Segall:2012.292} proposed the counter and value property in 2012, with the aim to reduce the complexity of representing some particular kinds of constraints in test models.

A counter is a special parameter, which can be used when constraints refer to the number of occurrences of parameter values. For example, assuming a SUT with ten parameters, each of which represents an operating system that can take values from \{{\em Windows}, {\em Linux}\}. A constraint may require that at least five operating systems should be {\em Windows}. This will result in $C(10, 6) = 210$ forbidden tuples. In this case, a counter $c$ can be defined to count the number of occurrences of {\em Windows} in a test case, so that only one constraint, i.e., $c \geq 5$, is required~\cite{Segall:2012.292}.

In addition, a value property can be associated with each parameter value to cater for constraints that are related to particular aspects of software parameters. For example, assume the operating system can take multiple {\em Windows} and {\em Linux} versions, such as \{{\em Windows 8}, {\em Windows 10}, {\em CentOS 7}, {\em Ubuntu 18}\}. 
A constraint may require that when a combination $a$ appears, the parameter $p$ must take a version of {\em Windows}. In this case, a platform property can be added to each parameter value, so that the constraint can be represented as `$a \rightarrow \textit{Platform}(p) = \textit{Windows}$'~\cite{Segall:2012.292}.

\subsubsection{Embedded Function}
In order to make constraint handling more flexible and accessible to software engineers, starting from 2015, Sherwood~\cite{Sherwood:2015.483, Sherwood:2016.563} proposed to represent constraints as embedded functions. 
The idea of this approach is to define constraints by a function embedded in the test model. This function is written in the programming language of the system, and will be used by a test generator to construct test cases.

For example, assuming that a SUT has three parameters {\em Year}, {\em Month} and {\em Day}, and {\em Day} will take values from the set \{1, 10, {\em last\_day}\}. Here, the value of {\em last\_day} is determined by the combination of {\em Year} and {\em Month}. In this case, a tester can write a function $f(year, month)$, which returns specific {\em last\_day}, as the third value of parameter {\em Day}~\cite{Sherwood:2015.483}.

\subsection{Impact of Constraints on CT}

The impact of constraints may vary with different problems, but in general, constraints increase the complexity and difficulty of effectively applying combinatorial testing:
\begin{enumerate}
\item Due to the lack of automated tools, it is usually a time consuming task to correctly identify the appropriate constraints for a given SUT. The manually extracted constraints may contradict each other, and some parts of constraints may be redundant. In addition, an under-constrained model will produce unexpectedly invalid test cases, while an over-constrained model will fail to examine a number of combinations that may trigger failures. 

\item It is hard to calculate the set of $\tau$-way combinations that need to be covered, which is a necessary component in many covering array generation algorithms. The number of parameters that are involved in a constraint may be greater than, equal to or less than $\tau$, and even a small number of constraints may produce a large number of invalid combinations. Moreover, implicit constraints may be introduced due to the interactions of other explicit constraints. Such implicit constraints are not obvious, so additional operations are usually required.

\item The size of a $\tau$-way covering array with constraints may become greater or less than its conflict free version. The presence of a constraint always reduces the number of all feasible test cases, but it does not guarantee the size reduction of the test suite needed to achieve $\tau$-way combination coverage. For example, if an orthogonal array (in which each $\tau$-way combination is covered exactly once) can be found for a SUT, then adding a forbidden tuple will increase the size of the required $\tau$-way covering array~\cite{Cohen:2007.114}. As a result, the lower and upper bounds of covering arrays may become unpredictable. This uncertainty limits the effective application of some generation algorithms that rely on such bounds to determine achievable size.

\item In order to support constraints, some constraint handling techniques should be integrated into covering array generation algorithms and tools. An inappropriate choice of constraint handler may dramatically decrease the performance of a generation algorithm. For example, Petke et al.~\cite{Petke:2015.512} found that the execution time of a greedy algorithm can be reduced from hours to seconds when using suitable techniques to handle constraints. However, the choices of either the best representation form of constraint, or the the best constraint handling technique, are still open problems.

\item Constraint violation may suffer from the {\em masking effect}~\cite{Yilmaz:2012.308}: if a test case fails to run due to an unsatisfied constraint, none of the valid combinations appearing in that test case will be tested. In particular, if a valid $\tau$-way combination is only covered by that failed test case, this will result in false confidence in the testing process as this combination is expected to be tested but, in fact, is not. For example, in Table~\ref{table_covering}, as the 2-way combination ($Italic$, $-$, $Red$, $-$, $-$) is only covered in an invalid test case $t_4$, this combination will not be tested by only executing those nine test cases.
\end{enumerate}

Although the presence of constraints brings a number of side effects, testers can build more accurate and flexible test models for the SUT with the help of constraints: they do not need to worry about introducing conflicts when determining parameters and values.
Moreover, a large number of constraints can greatly reduce the size of the search space, which makes the generation of high strength covering arrays feasible at a reasonable computational cost, especially for meta-heuristic search algorithms~\cite{Petke:2013.400, Petke:2015.512}. This is an important blessing that may be achieved from the presence of constraints, because there is evidence that testing higher strength interactions can lead to improved failure revelation~\cite{Kuhn:2004.59}.

\begin{table*}[!t]
\renewcommand{\arraystretch}{1.3}
\caption{References mapping research on constrained handling in combinatorial testing}
\label{table_reference_handle}
\centering
\begin{tabular}{l l p{11cm}}
\toprule
Category & Technique & References \\
\midrule
Remodel & Sub-model &~\cite{Sherwood:1994.10, Cohen:1997.14, Grindal:2006.98, Grindal:2007.102, Chen:2010.203, Othman:2011.248} \\
\cline{2-3}
& Abstract Parameter &~\cite{Williams:1996.13, Grindal:2007.102, Grindal:2006.98} \\
\hline
Avoid & Verify & \cite{Cohen:1997.14, Tung:2000.23, Czerwonka:2006.97, Grindal:2006.98, Grindal:2007.102, Lamancha:2010.236, Wang:2010.204, Alsewari:2012.309, Alsewari:2012.700, Calvagna:2012.343, Li:2012.315, Yu:2013.368, Othman:2014.470, Yu:2014.461, Ahmed:2014.871, Farchi:2014.421, Alsariera:2015.690, Lamancha:2015.544, Palaciosa:2015.501, Yu:2015.482, Gao:2016.632, Khalsa:2016.691, Khalsa:2018.874} \\
\cline{2-3}
& Solver & \cite{Cohen:2007.114, Cohen:2007.119, Cohen:2008.131, Garvin:2009.177, Yuan:2009.694, Garvin:2010.207, Oster:2010.235, Johansen:2012.328, Johansen:2012.328, Yu:2013.388, Haslinger:2013.399, Gao:2014.456, Hirasaki:2014.451, Lopez-Herrejon:2014.437, Jia:2015.526, Lin:2015.521, Al-Hajjaji:2016.698, Yamada:2016.577, Sheng:2017.699, Shwetha:2017.833, Bazargani:2018.757, Mercan:2018.852, Fogen:2018.820} \\
\cline{2-3}
& Weight & \cite{Bryce:2006.92, Bryce:2007.121} \\
\cline{2-3}
& Tolerate & \cite{Galinier:2017.676, Ahmed:2017.644} \\
\hline
Post & Replace & \cite{Hartman:2004.883, Grindal:2006.98, Grindal:2007.102, Huang:2010.226, Nakornburi:2016.628, Nakornburi:2016.692, Li:2016.713} \\
\cline{2-3}
& Expand & \cite{Yuan:2007.112, Yuan:2010.210} \\
\hline
Transfer & Constraint Satisfaction Problem & \cite{Hnich:2005.80, Hnich:2006.87, Calvagna:2008.124, Calvagna:2008.140, Calvagna:2009.189, Grieskamp:2009.175, Calvagna:2010.229, Perrouin:2010.225, Erdem:2011.237, Hervieu:2011.273, Johansen:2011.259, Nanba:2011.710, Kruse:2012.324, Nanba:2012.708, Yilmaz:2012.308, Gotlieb:2012.802, Henard:2013.346, Lopez-Herrejon:2013.387, Marijan:2013.345, Zhao:2013.353, Henard:2013.712, Henard:2014.445, Zhang:2014.468, Henard:2015.486, Sen:2015.506, Yamada:2015.529, Hervieu:2016.695, Mercan:2016.630, Liu:2016.702, Banbara:2017.723, Jin:2018.813} \\
\cline{2-3}
& Graph Model & \cite{Vilkomir:2008.136, Wang:2008.135, Danziger:2009.190, Maltais:2009.191, Vilkomir:2009.163, Maltais:2011.281, Salecker:2011.264, Segall:2011.276, Kruse:2012.325, Yu:2012.316, Gargantini:2014.458, Halle:2015.704, Duan:2017.650}\\
\bottomrule
\end{tabular}
\end{table*}

\section{Constraint Identification}
\label{sec_identification}

Modelling of the SUT is a fundamental activity in CT. One issue that needs to be resolved in modelling is to identify the constraints between parameters and their values. This has been recognised as an important component in modelling frameworks. For example, Grindal and Offutt~\cite{Grindal:2007.109} presented an eight step process for input parameter modelling, in which one step is to document invalid combinations.

Despite awareness that constraint identification is an indispensable part in modelling, considerable manual effort is often required, and there are few techniques that can be used to find and formalise constraints, especially by automated means.

One attempt to automatically identify constraints is to derive combinatorial test models from other analysis artefacts. Satish et al. proposed rule based techniques to semi-automatically extract the combinatorial test model from UML diagrams, including activity diagrams~\cite{Satish:2013.358}, sequence diagrams~\cite{Satish:2014.424} and use case diagrams~\cite{Satish:2017.657}. Their results for case analysis suggest that these techniques can partly reduce manual effort. However, support for constraint identification is either simply discussed without empirical evaluation, or left for future work.

Nyguyen and Tonella~\cite{Nguyen:2013.380} tried to apply machine learning technique to infer classifications and dependencies for CT. They assumed that inputs that execute similar code tend to belong to the same class, so that a clustering algorithm could be used to classify a set of randomly generated inputs into groups. An additional invariant inference was applied for each group, and the disjunction of invariants obtained from all groups thereby defines the dependencies that must hold. However, manual visual analysis was still needed in this approach and the usefulness of dependency inference was not evaluated.

Nakagawa and Tsuchiya proposed another two initial ideas to infer constraints. One technique~\cite{Nakagawa:2015.537} is based on the distance between words in requirement documents, which assumes that if two parameter values are always located near each other, they are more likely to cause constraints. A metric called ``coupling strength'' is proposed to measure the distance between two parameter values; high coupling strength pairs are used to define constraints.
Another technique~\cite{Nakagawa:2015.614} is based on a web search engine, which assumes that excessively high numbers of search hits indicate that the corresponding parameter values are highly related, whereas low numbers indicate these values are uncommon. A rate is calculated based on the search hits for any two parameter values, then constraints are defined from extremely low rate pairs, which should be excluded, as well as extremely high pairs, which should appear together.
However, these two techniques only focus on constraints between exactly two parameters, and only preliminary experiments are reported.

\begin{figure}[!t]
\centering
\includegraphics[width=0.5\textwidth]{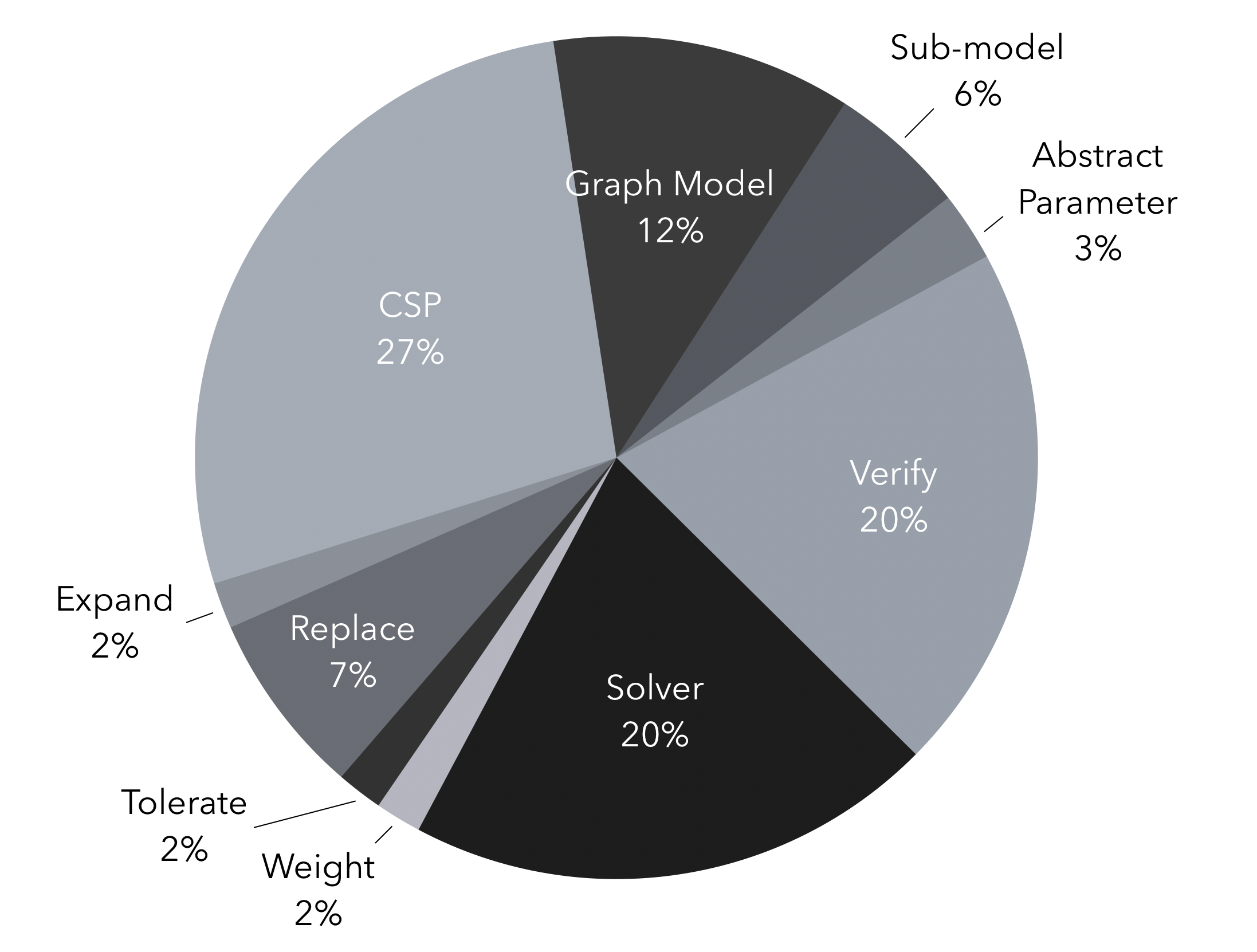}
\caption{The distribution of constraint handling techniques used in the surveyed studies.}
\label{fig_distribution_handling}
\end{figure}

\begin{figure*}[!t]
\centering
\includegraphics[width=\textwidth]{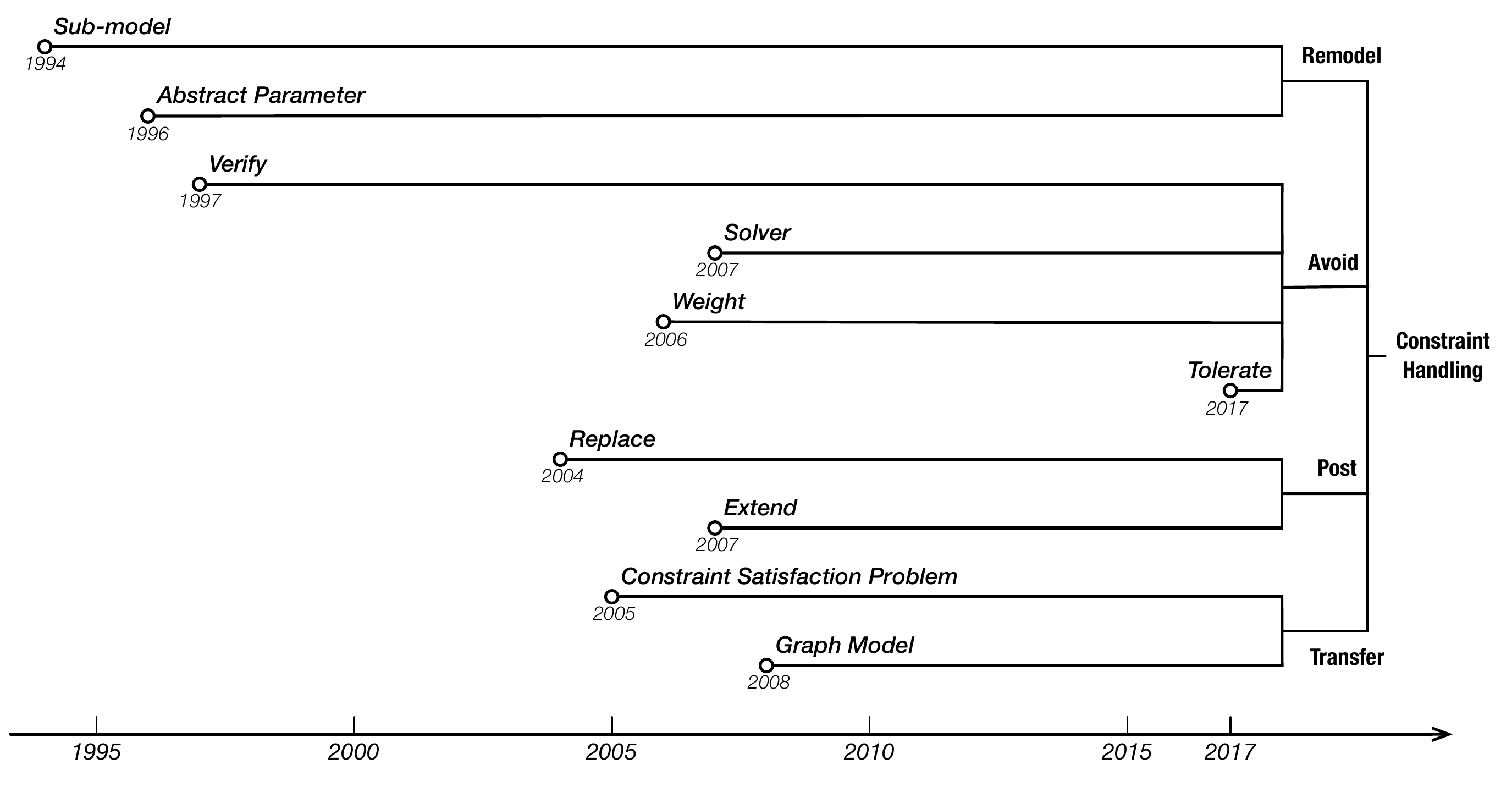}
\caption{The chronological development of constraint handling techniques.}
\label{fig_chronological}
\end{figure*}

\section{Constraint Handling}
\label{sec_handling}

Given a test model with constraints, generating a constrained covering array of the minimum size to cover all valid $\tau$-way combinations remains a challenging task in CT. Constraint handling focuses on this process, in order to ensure that the final covering array only contains valid test cases. This is the most prominent field in constrained combinatorial testing (see Figure~\ref{fig_distribution}).

In this study, we classified the available constraint handling techniques into four main categories:
\begin{enumerate}

\item {\bf Remodel}. Techniques to eliminate constraints through modifying a test model before test suite generation.

\item {\bf Avoid}. Techniques to construct conflict-free solutions through integrating particular strategies as extensions of existing algorithms during test suite generation.

\item {\bf Post-process}. Techniques to repair constraint violations through post-processing after test suite generation. 
 
\item {\bf Transfer}. Techniques to transfer the generation of constrained covering arrays into other problems, or use other structures to model the combinatorial input space, so that final solutions can be directly obtained by applying existing algorithms or tools.

\end{enumerate}
Moreover, we further classified sub-categories into specific techniques for each of the above categories. We will describe each of them in more detail in the following sections. We note that a few papers do not provide sufficient information to support a fully confident classification. They are nevertheless classified into the most likely category based on our understanding.

Table~\ref{table_reference_handle} provides references that map to each of the specific constraint handling techniques to each category. Note that multiple techniques can be used in one publication. 
Figure~\ref{fig_distribution_handling} shows the distribution of specific techniques used in the surveyed studies. We can see that {\em Avoid} and {\em Transfer} together account for 83\% of all studies, and the top three widely used specific techniques are {\em Constraint Satisfaction Problem}, {\em Verify} and {\em Solver}.

In addition, Figure~\ref{fig_chronological} gives an overview of the chronological development of these constraint handling techniques. We can see that {\em Remodel} is the earliest technique to be used. 
We can also see that most of these techniques appeared around 2006. It was the time of the first review of constraint handling techniques~\cite{Grindal:2006.98}. This was also the time of the first proposal of automated constraint handling techniques that do not need to either modify the model or explicitly list all forbidden combinations~\cite{Cohen:2007.114}. Recently, even though many techniques were already available, a new technique, {\em Tolerance}, was proposed in 2017~\cite{Galinier:2017.676, Ahmed:2017.644}, suggesting that the field remains open for innovation.


\subsection{Remodel}

The `remodel' technique focuses on constraint handling before test suite generation. Its aim is to eliminate constraints from test models, so that conventional covering arrays can be used directly. We describe two specific techniques for this category: {\em Sub-model} and {\em Abstract Parameter}.

\subsubsection{Sub-model}
The `sub-model' technique removes constraints by constructing a set of conflict-free sub-models. Test suites are generated separately for each sub-model and then combined later. This idea was initially sketched in 1994, where Sherwood presented an approach of listing allowed value combinations into different groups for the Constrained Array Test System (CATS)~\cite{Sherwood:1994.10}. Later this approach was also used in the AETG system~\cite{Cohen:1997.14}. Additionally, Chen et al.~\cite{Chen:2010.203}, and Othman and Zamil~\cite{Othman:2011.248} used similar ideas to handle shielding constraints.

\begin{figure}[!t]
\centering
\includegraphics[width=0.45\textwidth]{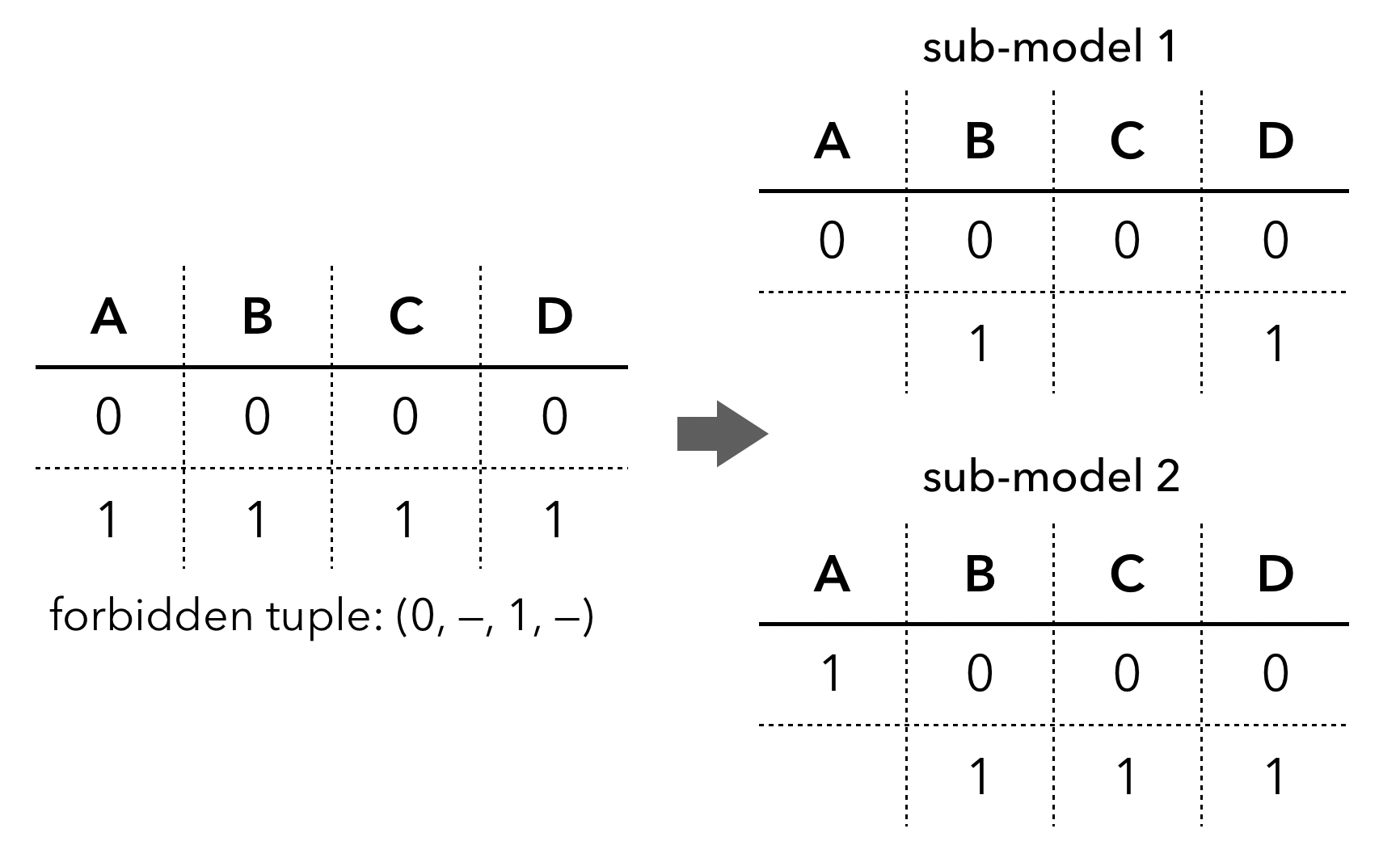}
\caption{An example of sub-model technique.}
\label{fig_submodel}
\end{figure}

To remove constraints, the sub-model approach firstly selects a split parameter, namely a parameter that is involved in a constraint with the least number of values. Secondly, the model splits into multiple intermediate sub-models, each of which contains one value of the split parameter and all values of other parameters. Then, for each intermediate sub-model, the values of other parameters that violate constraints are further removed. This process is repeated until no constraint remains, and finally the resulting constraint-free sub-models are merged to reduce the number of models. Figure~\ref{fig_submodel} shows an example of splitting a model into two conflict-free sub-models, where $A$ is selected as the split parameter.

\subsubsection{Abstract Parameter}
The `abstract parameter' technique removes constraints by combining conflicting parameters into abstract parameters. Then a test suite is generated for the new model and transformed back to the values of original parameters. This idea was firstly applied in 1996, when Williams and Probert discussed how it can be used with orthogonal array construction~\cite{Williams:1996.13}.

\begin{figure}[!t]
\centering
\includegraphics[width=0.45\textwidth]{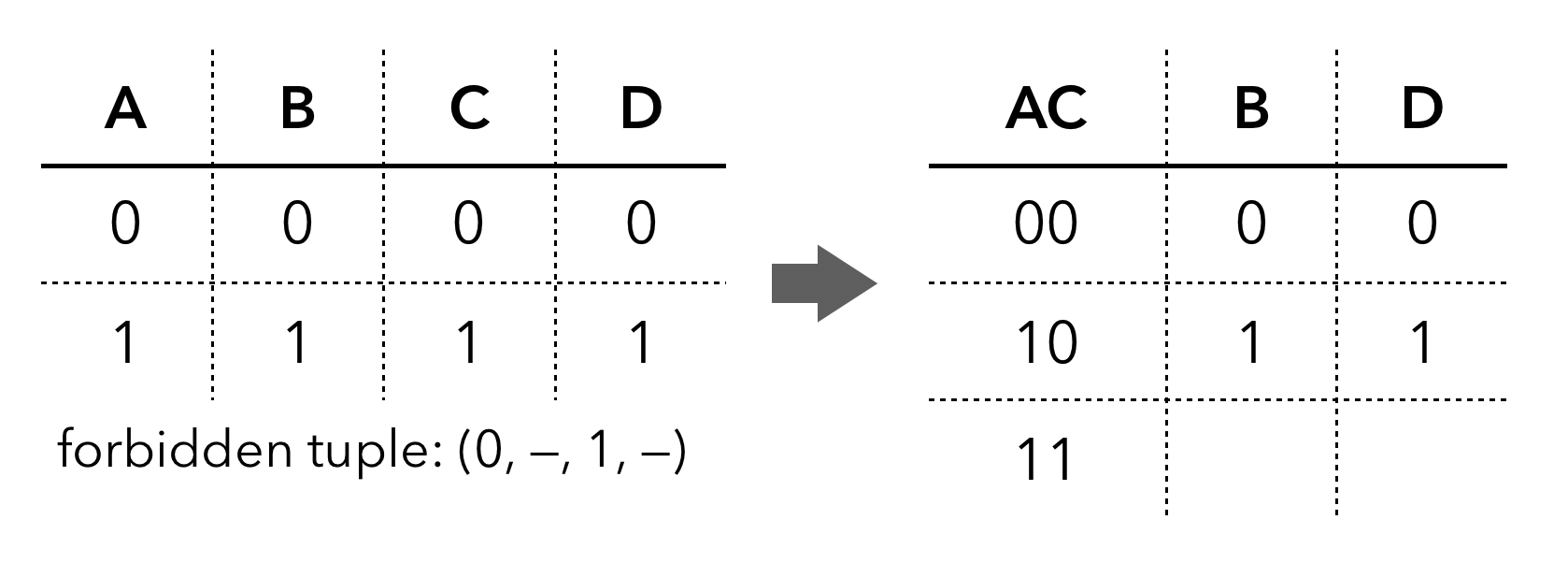}
\caption{An example of abstract parameter technique.}
\label{fig_abstract}
\end{figure}

To remove constraints, the abstract parameter technique firstly identifies conflicting parameters that are involved in constraints. Then these parameters are replaced by one or more abstract parameters whose values represent all valid sub-combinations. Figure~\ref{fig_abstract} shows an example of replacing two conflicting parameters $A$ and $C$ by an abstract parameter $AC$.

The sub-model and abstract parameter techniques are the two earliest techniques to handle constraints in CT. They can work well with a few simple constraints, but they are less competitive in the size of generated test suite~\cite{Grindal:2006.98, Grindal:2007.102}. Moreover, these techniques usually need manual effort to modify the test model, which makes them difficult to scale to large test models with complex constraints~\cite{Cohen:2007.114}.


\subsection{Avoid}

To generate a constrained $\tau$-way covering array, three main frameworks can be used: (1) constructing a single test case that maximises coverage at each time (one-test-at-a-time); (2) firstly constructing a test set for the first $\tau$ parameters and then extending it horizontally and vertically (in-parameter-order); and (3) directly constructing a covering array for a given size (evolve-test-suite).
These three frameworks can be combined with both greedy and search based strategies. In either case, the generation strategy will apply a step-by-step process by assigning or changing parameter values, in order to cover as many combinations as possible. Finally, a covering array is achieved when all valid $\tau$-way combinations are covered.

The `avoid' technique focuses on constraint handling during the above generation process. Note that this technique is usually used as an extension of a covering array generation algorithm. This differs from some techniques that directly produce valid final solutions. We describe four specific techniques for this category: {\em Verify}, {\em Solver}, {\em Weight}, and {\em Tolerance}.

\subsubsection{Verify}

The `verify' technique is probably the most basic technique to handle constraints in the more general case. The idea is to maintain a list of forbidden tuples, and each partial or complete solution during the generation process will be verified against them to prevent the appearance of constraint violation. The AETG system~\cite{Cohen:1997.14} first used this technique to determine disallowed test cases.

One important aspect of this technique is that all forbidden tuples must be explicitly listed in advance and retained in memory. Some algorithms need to pre-construct such a list~\cite{Alsewari:2012.309, Alsewari:2012.700, Lamancha:2015.544}, but it may be error-prone and impractical due to implicit constraints (as discussed in Section~\ref{sec_definition}).
In order to automatically deduce implicit constraints, Li et al.~\cite{Li:2012.315} proposed a method based on the observation that implicit constraints are introduced when every value of a parameter is involved in some forbidden tuples. Later, similar idea was also used in Yu et al.'s studies~\cite{Yu:2014.461, Yu:2015.482}.

Another issue of this technique is the need to check a large number of forbidden tuples. To quickly verify the validity, Yu et al.~\cite{Yu:2014.461} proposed a notion named {\em Minimum Forbidden Tuple} (MFT). An MFT is a forbidden tuple of minimum size that covers no other forbidden tuples. Once all MFTs are found, validity verification can be performed by only checking whether a solution contains any MFT. To generate all MFTs, they proposed an algorithm to iteratively derive new (implicit) forbidden tuples and to simplify the current set of forbidden tuples until no new tuples can be derived.
Moreover, not all MFTs are needed when verifying a partial test case that only relates to a subset of parameters. Based on this observation, Yu et al.~\cite{Yu:2015.482} further proposed an on-demand MFT generation strategy to only derive necessary MFTs. This strategy utilises the fixed values of partial test cases to remove values from forbidden tuples. For example, given a partial test case ($1$, $1$, $-$, $-$), the forbidden tuple ($-$, $1$, $0$, $-$) can be simplified to ($-$, $-$, $0$, $-$).

\subsubsection{Solver}
\label{sec_solver}

Instead of verifying against a list of forbidden tuples, the `solver' technique aims to encode constraints and solutions into a formula and apply an existing constraint satisfaction solver to check the formula's validity. Cohen et al.~\cite{Cohen:2007.114} first suggested this idea as a general solution to resolve constraints in 2007. They integrated a SAT solver into both greedy and simulated annealing based covering array generation algorithms to ensure that each value assignment or change is constraint satisfying. This first attempt later gave rise to two popular algorithms: the SAT-based AETG approach~\cite{Cohen:2008.131} and the hill climber CASA~\cite{Garvin:2009.177, Garvin:2010.207}.

One of the most widely used constraint solvers in this technique is the SAT solver. To do this, the constraints should be encoded into a boolean formula. A common approach is to introduce a boolean variable $x_{ij}$ for each $p_i \in P$ and $v_j \in V_i$, where $x_{ij}$ is true represents that parameter $p_i$ takes value $v_j$ (i.e., $p_i = v_j$).
A constraint $a$ can thus be encoded as a boolean formula $\phi_a$ over these boolean variables. For example, a $k$-way forbidden tuple $a$ can be encoded as $\phi_a = \neg \bigwedge_{x_{ij} \in Q} x_{ij}$, where $Q$ is the set of fixed parameter value assignments in the forbidden tuple ($|Q|=k$). 
Then the set of all constraints $F$ in the test model can be encoded as:
$$C_F = \bigwedge_{a \in F} \phi_a$$
Additionally, each parameter should have one and exactly one value in a test case. This can be encoded by an at-least constraint~\cite{Cohen:2008.131}:
$$C_L = \bigwedge_{p_i \in P}(\bigvee_{v_j \in V_i} p_i = v_j)$$
and an at-most constraint~\cite{Cohen:2008.131}:
$$C_M = \bigwedge_{p_i \in P}(\bigwedge_{v_j, v_j' \in V_i, v_j \neq v_j'} p_i \neq v_j \vee p_i \neq v_j')$$
Together, $C = C_F \wedge C_L \wedge C_M$ represents the base constraint~\cite{Cohen:2008.131}. Now given the encoding of any partial or complete test case $S$, a SAT solver can be used to decide whether it is possible to find a truth assignment for the variables in the boolean formula $C \wedge S$ (namely, whether $S$ is constraint satisfying).
Here implicit constraints are accounted in the constraint solving process, so that they do not need to be explicitly included in $C_F$.

When encountering an invalid intermediate solution during the generation, one approach is to simply abandon this solution and try the next best candidate~\cite{Cohen:2008.131, Garvin:2010.207, Yu:2013.388, Lin:2015.521}. Another approach is to repair the solution, such as randomly changing parameter values, until it becomes constraint satisfying~\cite{Jia:2015.526, Yamada:2016.577}.

Verifying validity after each change usually leads to a large number of solver calls. One simple optimisation is to use the solver only when the parameters of the changed values are mentioned in constraints~\cite{Cohen:2007.114, Yu:2013.388, Gao:2014.456}. In addition, some characteristics of the solver can also be used to improve efficiency. Cohen et al.~\cite{Cohen:2007.119, Cohen:2008.131} mined the solving history to prune the search space for AETG, so that only a fraction of value assignments need be examined. Yu et al.~\cite{Yu:2013.388} and Bazargani et al.~\cite{Bazargani:2018.757} recorded previous solving results to avoid multiple validity verifications for the same test. Recently, the unsatisfiable core, which can be used to find forbidden tuples in an unsatisfiable test, was used by Yamada et al.~\cite{Yamada:2016.577}. It enabled a lazy detection of invalid combinations so that they do not need to be removed in the pre-processing step. They also proposed an `amend' method to repair invalid test cases by removing value assignments in the unsatisfiable core. Lastly they extended a SAT solver to support approximate answers to further reduce the number of solver calls while maintaining a low chance of producing invalid solutions.

\subsubsection{Weight}

Assigning weights to parameter values is a common method in test suite prioritisation~\cite{Choi:2015.519, Wu:2016.590}. Bryce and Colbourne~\cite{Bryce:2006.92, Bryce:2007.121} used this idea to cater for soft constraints. In their method, combinations were weighted as either important with positive values or undesirable with negative values. The Deterministic Density Algorithm (DDA)~\cite{Colbourn:2004.58} was then extended to generate a prioritised test suite to avoid those undesirable combinations where possible, so that when they are included they will only appear in the last tests of the test suite. This technique did not guarantee the exclusion of constraints; the undesirable combinations occurred around 3\% of the time~\cite{Bryce:2006.92}.

\subsubsection{Tolerate}\label{section_tolerate}

When using search-based algorithms to generate constrained covering arrays, one approach is to exclude all invalid solutions from the search space, so that all intermediate and final solutions are constraint satisfying. However, sometimes some elements of invalid intermediate solutions may help to find the optimal solution, so it may be desirable to `tolerate' such solutions during the generation.

Incorporating a penalty term into the fitness function is one of the most common approaches of solving constrained optimisation problems by search-based algorithms~\cite{Coello:2002}. It includes invalid solutions in the search space, but penalises them in favour of valid solutions. 
More recently, Galinier et al.~\cite{Galinier:2017.676} used this technique with a Tabu search algorithm to generate constrained covering arrays. They encoded the whole test suite as a candidate solution, and used the following fitness function to evaluate the goodness of each candidate $s$:
$$f(s) = U(s) + \omega \cdot V(s)$$
where $U(s)$ is the number of yet to be uncovered combinations, $V(s)$ is the number of constraint violations in $s$, and $\omega$ is the penalty weight. 
Here, the value of $V(s)$ is easy to calculate if there are only a few constraints. When the set of constraints is large and complex, a more efficient approach, such as the Minimal Forbidden Tuple (MFT) approach or a constraint satisfaction solver, should be used to determine whether a constraint is violated in the solution $s$.

With the above fitness function, the search will stop when $f(s)=0$ is met, i.e., a constrained covering array is found, or a maximum number of iterations is reached. As the fitness function is an indispensable part of any search-based algorithm, this technique can be directly used in other search-based test suite generation algorithms.

In addition, Ahmed et al.~\cite{Ahmed:2017.644} also used similar idea to design a new particle swarm optimisation algorithm to generate constrained covering arrays. Their approach generates each test case one at a time, with two objectives in mind: the first objective is to contain as few constraint violations as possible, and the second objective is to cover as many of the yet uncovered combinations as possible. These two objectives are applied in order to evaluate each candidate test case. The best test case is the one with maximum combination coverage and no constraint violation.


\subsection{Post-process}

The `post-process' technique focuses on constraint handling after test suite generation. Given a covering array generated without considering constraints, this technique first finds all invalid test cases, which can be efficiently achieved by using the Minimal Forbidden Tuple (MFT) approach or a constraint satisfaction solver, as introduced before.
Then the conflicts in these invalid test cases can be removed by different means. We describe two specific techniques for this category: {\em Replace} and {\em Extend}.

\subsubsection{Replace}\label{section_replace}

The `replace' technique resolves conflicts by replacing invalid test cases by a set of valid ones to remove constraint violations while retaining combination coverage. Hartman and Raskin~\cite{Hartman:2004.883} first used this technique to design Combinatorial Test Services (CTS) package, in which they did not implement an algorithm to construct constrained covering arrays, but provided a service to `delete and perturb' columns that contain forbidden tuples.

Given an invalid test case, a straightforward approach of replacing it is to clone this test case. In each clone, the parameters involved in constraints are changed to new values that do not violate constraints, and the number of clones is chosen to preserve coverage~\cite{Grindal:2006.98, Grindal:2007.102, Nakornburi:2016.628, Nakornburi:2016.692, Li:2016.713}. 
Another attempt is to first remove invalid test cases and then use a genetic algorithm to add some feasible ones~\cite{Huang:2010.226}.

\subsubsection{Extend}

The `extend' technique aims to directly extend invalid test cases to remove conflicts. Yuan et al.~\cite{Yuan:2007.112, Yuan:2010.210} used this technique to handle constraints in event sequences. In their approach, abstract test sequences were first generated. As some events can only be executed with some prior set-up events, new events were added into test sequences to create the final executable test cases. For example, in GUI testing, the {\em Undo} will be disabled unless some specific events occur. A {\em Copy} event can thus be inserted before {\em Undo} to produce an executable sequence~\cite{Yuan:2010.210}.


\subsection{Transfer}
The `transfer' technique focuses on reformulating the problem of covering array generation into other problems, or using other structures to model the constrained input space. We describe two techniques for this category: {\em Constraint Satisfaction Problem} and {\em Graph Model}.

\subsubsection{Constraint Satisfaction Problem (CSP)}
When handling constraints by the `avoid' technique, a constraint solver is only used to determine whether a partial or complete solution is constraint satisfying. The same process can also be used to produce satisfiable test suites or test cases: the covering array generation problem is reduced to the CSP problem and then existing solvers can be directly applied. This idea of problem reduction can be traced back to 2005, when Hnich et al.~\cite{Hnich:2005.80, Hnich:2006.87} encoded the whole covering array into a boolean formula and used a SAT solver to find feasible solutions. They mentioned that constraints can be easily handled, but this was not implemented nor evaluated. Later similar ideas were used by Calvagna et al.~\cite{Calvagna:2008.140} and Grieskamp et al.~\cite{Grieskamp:2009.175} to generate constrained covering arrays.

The constrained covering array generation problem can be encoded to different constraint satisfaction problems. For example, in order to use a SAT solver to generate a test suite, one approach is to use the matrix encoding~\cite{Hnich:2005.80, Yamada:2015.529}.
Assuming that the test suite is represented by an $m \times n$ matrix. Let a boolean variable $x_{ijv} (1 \leq i \leq m, 1 \leq j \leq n, v \in V_j)$ denote the fact that the value of the $j$-th parameter is $v$ in the $i$-th row. Here, each parameter must have a unique value in each row, which can be encoded by at-least and at-most constraints (as in Section~\ref{sec_solver}), or encoded by a uniqueness constraint~\cite{Yamada:2015.529}:
$$U = \bigwedge_{1 \leq i \leq m} \bigwedge_{1 \leq j \leq n}(1 = \sum_{v \in V_j} x_{ijv})$$
To ensure the requirement of $\tau$-way coverage, let $O$ be the set of all valid $\tau$-way combinations to be covered, and a boolean variable $c_{ik}$ denote that combination $k$ ($k \in O$) is covered in the $i$-th row. 
The coverage constraint can thus be encoded as follows~\cite{Yamada:2015.529}:
$$C_1 = \bigwedge_{k \in O} \bigvee_{1 \leq i \leq m}c_{ik}$$
Additionally, to ensure the consistency between $x_{ijv}$ and $c_{ik}$, the following constraints should be satisfied~\cite{Yamada:2015.529}:
$$C_2 = \bigwedge_{1 \leq i \leq m} \bigwedge_{k \in O} \bigwedge_{j, v \in k}(c_{ik} \Rightarrow x_{ijv})$$
Every test case should also satisfy the set of constraints $F$ in the test model. For each row $i$, a constraint $a \in F$ can be encoded into a boolean formula $\phi_{ia}$ over variables $x_{ijv}$. This results in the following constraint~\cite{Yamada:2015.529}:
$$C_3 = \bigwedge_{1 \leq i \leq m} \bigwedge_{a \in F} \phi_{ia}$$
Lastly, a SAT solver is invoked to ask for the solution to the contained form $U \wedge C_1 \wedge C_2 \wedge C_3$. If there is a satisfiable assignment, we find a $\tau$-way constrained covering array of size $m$.

Apart from encoding a test suite as SAT constraints, other encodings and solvers have also been used. Calvagna and Gargantini~\cite{Calvagna:2008.124, Calvagna:2008.140, Calvagna:2010.229} used test predicates to formalise the constrained covering array generation problem, and applied model checker to construct each test case at a time. Their approach can deal with temporal constraints, where parameter values may change over time~\cite{Calvagna:2008.124}. A Satisfiability Modulo Theory (SMT) solver was also used~\cite{Calvagna:2009.189, Grieskamp:2009.175}. It uses a smaller number of variables than the boolean encoding, but Henard et al.~\cite{Henard:2015.486} found that the processing time required by an SMT solver was higher than that required by the SAT solver.

In addition, some methods try to find solutions that not only satisfy constraints but also optimise some objectives, namely to solve an optimisation problem instead of a pure satisfiability problem. Constraint programming~\cite{Hervieu:2011.273, Hervieu:2016.695} and linear programming~\cite{Kruse:2012.324, Lopez-Herrejon:2013.387} were both used to generate the smallest test suites, and pseudo-Boolean optimisation~\cite{Zhang:2014.468} was used to generate each test case that maximises current combination coverage.


CSP-based techniques can generate the covering array of minimum size and prove its optimality, but it usually suffers from the scalability problem. One problem is that the number of variables and clauses increases dramatically with the increase of test model and covering strength. Perrouin et al.~\cite{Perrouin:2010.225} proposed two `divide-and-conquer' strategies to divide constrained covering array generation into small and solvable problems and then combine sub-solutions. 
Another problem is that many solver runs are usually required to determine the minimum size of a covering array. In this line, Nanba et al.~\cite{Nanba:2011.710, Nanba:2012.708} used binary search; Yamada et al.~\cite{Yamada:2015.529} used incremental SAT solving, where the learned clauses for solving the problem of size $m$ can be reused when solving the problem of size $m - 1$.
Nevertheless, it is usually reported that the CSP-based techniques have high computational cost, and they are only practicable for pairwise testing~\cite{Nanba:2011.710, Nanba:2012.708, Zhang:2014.468, Yamada:2015.529}.

\subsubsection{Graph Model}
Apart from modelling input spaces using a list of parameters and their values, another technique is to use a graph structure. This allows graph-related operations and theories to be used directly to construct test cases or test suites. 

The application of such graph-related techniques first appeared in 2008~\cite{Wang:2008.135, Vilkomir:2008.136}, where graph structures were used to build conflict-free models. Wang et al.~\cite{Wang:2008.135} used a navigation graph to model web applications. In their approach, each node represents a web page, and each edge represents a possible link from one node to another. Constraints are implicitly captured with this structure: each path through the graph is a valid test sequence, and a set of paths is generated to cover all ordered node pairs. Similar techniques were also proposed by Vilkomir et al.~\cite{Vilkomir:2008.136, Vilkomir:2009.163}, where a Markov chain model was used. Later, Yu et al.~\cite{Yu:2012.316} investigated efficient algorithms for generating test sequences. They used a labeled transition system graph as the model, which consists of states, event labels and transitions between states.

Another application is to use a Binary Decision Diagram (BDD). A BDD is a structure that represents and manipulates boolean formulas: each non-terminal node represents a variable, outgoing edges from a non-terminal node represent values of corresponding variables, and a path from the root to a terminal node represents an assignment, which evaluates to true if it ends in 1-terminal node or false otherwise~\cite{Salecker:2011.264}. 
Salecker et al.~\cite{Salecker:2011.264} modelled the space of all valid test cases as a single boolean formula represented by a BDD. Test cases were then generated to be compatible with this BDD and to maximise combination coverage. Gargntini and Vavassori~\cite{Gargantini:2014.458} proposed a similar approach, but they used a Multivalued Decision Digram (MDD) instead of BDD. Segall et al.~\cite{Segall:2011.276} also used a BDD to capture the uncovered combinations for each $k$ parameters. Then test cases were generated relying on conjunction and counting satisfying assignment operations of BDDs.

After modelling using graph structures, test suites can also be generated by applying graph-related theories. Danziger et al.~\cite{Danziger:2009.190} used a covering array with forbidden edge (CAFE) to generate test suites, which is closely related to the edge clique covering problem~\cite{eccp:1968}. Maltais et al.~\cite{Maltais:2009.191, Maltais:2011.281} further investigated the computational complexity of determination and approximation of CAFEs. Recently, algorithms for graph colouring were also used for generating the smallest covering array~\cite{Halle:2015.704}, or optimising the vertical growth of IPO algorithm~\cite{Duan:2017.650}.


\section{Constraint Maintenance}
\label{sec_maintenance}

Modelling of combinatorial testing is usually a tedious and error-prone task. In practice we cannot assume either the creation of a perfect test model at one time, or the usage of a model that never changes with the evolution of SUT. Particular techniques are needed to comprehend and to manage test models, which include validating, repairing and evolving constraints in the test model.

Model review is a basic technique to manually validate test models. Farchi et al.~\cite{Farchi:2013.364} proposed to repeatedly review projections of the test space on a subset of parameters in order to find and debug modelling mistakes. For a reviewed projection, each value combination is determined as valid, invalid or partially valid based on BDD and SAT solving, so that testers can verify that these statuses are as they expected. Implicit constraints are also identified and displayed to attract special attention.
In addition, Tzoref-Brill et al.~\cite{Tzoref-Brill:2016.578} presented three visualisation forms, including matrix, graph and treemap, to visualise both test model and test suite. One of their goals is to display the relationships between parameters, constraints and parameter combinations in the model in an informative way, such that testers can easily confirm understanding of requirements and further tune the model.

One weak aspect of model review is that it is usually done by hand, even if it can be assisted by tools. In order to reduce manual effort for validation, Arcaini et al.~\cite{Arcaini:2014.431} proposed to automatically check some properties that any test model and test suite should hold: for a test model, the set of constraints should be consistent and does not contain redundant parts; there are no parameters or values that are never taken due to constraints. Their approach was implemented using an SMT solver, and they further provided strategies to deal with situations that violate the above properties.

In addition to validating the properties of the test model, another aspect is to validate the conformance between specification (i.e., test model) and implementation (i.e., the actual program). Gargantini et al.~\cite{Gargantini:2016.615} devised four policies to generate test cases to check whether the constraints in the test model correctly capture the relationships among parameters. They assumed that there exists an oracle function to determine the validity of a test case against the implementation. Then if the validity of a test case as defined in the model is different from actual validity, a conformance fault (either in the model or in the program) is found.
Recently, Gargantini et al.~\cite{Gargantini:2017.659} further proposed techniques to automatically repair conformance faults. A test case is marked as ``failed'' if it produces different validity results, or marked as passed otherwise. Then a combinatorial fault diagnosis tool was used to identify a set of suspicious combinations that cause the failing result, according to which constraints can be modified to fix the original test model.

Moreover, the test model will barely stay unchanged in practice, especially when the SUT evolves due to iterative development and bug fixing. In this case, additional efforts are often required to address the changes on the test model in order to maintain validity. 
Tzoref-Brill and Maoz~\cite{Tzoref-Brill:2015.531} found that the widely used boolean semantic is inadequate for model evolution. For example, assuming two parameters $p_1$ with $V_1 = \{0, 1, 2\}$ and $p_2$ with $V_2 = \{0, 1\}$. A constraint is represented as either $p_1 = 0 \rightarrow p_2 = 0$ or $p_1 = 0 \rightarrow p_2 \neq 1$, either of which produces the same input space. However, when a new value $2$ is added into $V_2$, the combination $(0, 2)$ will become invalid according to the former representation, but valid according to the latter representation~\cite{Tzoref-Brill:2015.531}. 

To resolve this problem, Tzoref-Brill and Maoz~\cite{Tzoref-Brill:2015.531} proposed to use lattice-based semantics that provide consistent interpretations and expose additional changes after atomic operations. Spichkova et al.~\cite{Spichkova:2016.619} proposed a human-centric method that asks the tester to make additional decisions based on a visual representation.
Recently, Tzoref-Brill and Maoz~\cite{Tzoref-Brill:2017.680} presented a canonical representation of test model, and developed algorithms to calculate and present both syntactic differences (i.e. addition and removal of parameters, values and constraints) and semantic differences (i.e. the set of valid test cases) between test models. Their aim is to assist testers in their need of verifying the completeness and correctness of model updates.

\section{Conclusion}
\label{sec_conclusion}

In this survey, we provide an overall picture of research work on constraints in Combinatorial Testing (CT). Based on the 129 papers published between 1987 and 2018, we summarised the different representations used to capture constraints, and discussed the influence of constraints on the successful applications of CT. 
We classified the studies into three overall research categories: constraint identification, constraint handling and constraint maintenance. The currently available techniques for each of the categories were then reviewed and analysed.

Although a rich collection of constraints pertinent techniques have been developed and applied in CT, as this study reveals, the issue of handling constraints remains challenging with many open problems. Specifically, there is a need to develop more powerful and automated algorithms for resolving constraints, and also to compare their performance against each other. The scarce researches on constraints identification and maintenance also call for further studies in these potentially important fields.


%

\section*{Acknowledgments}

This work was partially supported by the National Key Research and Development Plan (No. 2018YFB1003800). This work was also partially supported by the DAASE EPSRC Grant (No. EP/J017515/1) and EPSRC Fellowship (No. EP/P023991/1).

\ifCLASSOPTIONcaptionsoff
  \newpage
\fi



\bibliographystyle{IEEEtran}
\bibliography{manuscript-survey}

\end{document}